\def\ignorethis#1{}
\font\zcmr=cmr6
\font\ztitle=cmr10 scaled\magstep3
\def\f1t{\textfont1=\scriptscriptfont1}
\def\fs{\scriptfont0=\zcmr}
\def\f0ss{\scriptfont0=\scriptscriptfont0}
\baselineskip=\normalbaselineskip

\def\newpage{\vfill\eject}

\newcount\equationnumber
\newcount\sectionnumber
\sectionnumber = 0
\def\theEquationnumber{\the\sectionnumber.\the\equationnumber}
\def\zp{%
  \global\advance\equationnumber by 1
  (\theEquationnumber)}
\def\zpadvnoprint{%
  \global\advance\equationnumber by 1}
\def\zpstay#1{(\theEquationnumber{#1})}
\def\advsectionnumber{%
  \equationnumber=0 %
  \global\advance\sectionnumber by 1}

\def\sln{{sl}_{\rm n}}
\ignorethis{

 }

\def\ha{H_0^{\rm atom}}
\def\hp{H_0^\phi}
\def\hi{H_{\rm int}}

\def\dt{\partial_t}
\def\dx{\partial_x}
\def\er#1{| {\rm #1} \rangle}
\def\el#1{\langle {\rm #1} |}
\def\w#1#2{\omega_{#1#2}}
\def\wm#1{\omega_{#1}}
\def\wt#1#2{{\widetilde \omega}_{#1#2}}
\def\en#1{\epsilon_{#1}}
\def\dl#1#2{\delta_{#1#2}}
\def\s#1#2#3{\sum_{#1=#2}^{#3}}
\def\ba{b^\dagger}
\def\o#1#2{{\cal O}_{#1#2}}
\def\h#1{{\cal H}_#1}
\def\c#1{A^{-1}_{#1}}
\def\p{\vec p}
\def\x{\vec x}
\def\a{\vec A}
\def\ax#1{ {\vec A} ({\vec #1}) }
\def\d{\vec d}
\def\al#1#2{\alpha_{#1#2}}
\def\n{\widehat n}
\def\b#1#2{\beta_{#1#2}}
\def\bm#1{\beta_#1}
\def\ps#1#2{\psi_{#1#2}}
\def\pa#1#2{\psi^{\dagger}_{#1#2}}
\def\hf{}
\ignorethis{\def\hf{ {1\over 2\sqrt{2} } }}

\headline={\hfil CLNS 97/1472; hep-th/9703212}
\footline={3/97\hfil}

\topskip=1in
\centerline{{\ztitle A 1D Model for N-Level Atoms Coupled to an EM Field}}
\vskip1in
\centerline{Zorawar S. Bassi}
\centerline{Andr\'e LeClair}
\medskip
\centerline{Newman Laboratory}
\centerline{Cornell University}
\centerline{Ithaca, NY 14853}
\vskip0.8in

We construct a model for n-level atoms coupled to quantized 
electromagnetic fields in a fibrillar geometry.  In the slowly varying
envelope and rotating wave approximations, the equations of motion are 
shown to satisfy a zero curvature representation, implying integrability
of the quantum system.
\newpage

\topskip=10pt
\advance\pageno by -1
\headline={\hfil}
\footline={\hfil\folio\hfil}
\noindent {\bf Introduction}
\medskip

The interaction of radiation with two-level atoms has been extensively
studied under various approximations.  In one spacial dimension, the reduced
Maxwell-Bloch equations resulting from the slowly varying envelope and
rotating wave approximations are known to be quantum integrable [1].  In 
this paper we generalize the one-dimensional case of two-level atoms to that
of n-level atoms.  

In the first of two parts, we construct the fully quantum n-level model.  
The system consists of n-level atoms distributed in a fibrillar geometry, 
interacting with radiation through a minimally coupled hamiltonian.  In the
remaining section we apply the approximations, and show that the Heisenberg 
equations of motion for the reduced system satisfy the so-called 
zero curvature representation.  This implies that the system is integrable
and can be solved by the inverse scattering method.

\bigskip
\noindent {\bf \the\sectionnumber. Mathematical Background}
\medskip

Let us first recall the $\sln$ Lie algebra.  The $n^2 - 1$ generators,
written as
$$ \bigl\{ E_{ij}, H_m\ |\, 1\leq i \neq j \leq n, 1\leq m \leq r \bigr\},
\eqno\zp $$
where $r = n-1$ is the rank, satisfy the following brackets 
(in the Chevalley basis)
\zpadvnoprint
$$\fs \eqalignno{ \bigl[E_{ij}, E_{kl} \bigr] &= 
\left\{ 
\eqalign{ \dl kj E_{il} - \dl il E_{kj}, & 
{\rm \ if \ } \dl il \dl kj = 0 \cr 
\sum_{m=i}^{j-1} H_m \ (i<j), & 
{\rm \ if \ } \dl il \dl kj = 1 \cr } 
\right\}, & \zpstay{\rm a} \cr
\bigl[ E_{ij}, H_m \bigr] &= \bigl( \dl jm - \dl im - \dl j{m+1} +
\dl i{m+1} \bigr) E_{ij}, & \zpstay{\rm b} \cr
\bigl[ H_a,H_b \bigr] &= 0. & \zpstay{\rm c} \cr  } $$
The set spanned by $\bigl\{ H_m \bigr\}$ being the Cartan subalgebra.
(Note that in the bracket (0.2a) a term of the form $\dl i{\neq j} E_{kk}$ is
formally equal to zero, even though $E_{kk}$ has not been defined.  The
set $\bigl\{E_{ij}\ |\, 1\leq i,j\leq n\bigr\}$ satisfying the first
relation in (0.2a) is a basis for the algebra $gl_{\rm n}$.)
A representation $\rho$ of $\sln$ will be denoted as
$$\bigr\{E_{ij}^{\rho} = \rho(E_{ij}),H_m^{\rho} = \rho(H_m)\bigr\}.
\eqno\zp$$
The $r\times r$ Cartan matrix $A$ has the explicit form
$$\fs A_{uv} = 2\dl uv - \dl u{v-1} - \dl u{v+1}. \eqno\zp $$
It is a symmetric matrix with diagonal elements 2 and nearest off-diagonal
elements $-1$.  We now proceed to build our quantum system.

\bigskip
\advsectionnumber
\noindent {\bf \the\sectionnumber. The Interacting N-Level Hamiltonian}
\medskip

We model a free n-level atom as having a single electron with 
eigenstates $\er i, {\rm i} = 1,2,\ldots,{\rm n},$ and energies
$ \en 1 > \en 2 > \ldots > \en n$.  
The energy splitting between states $\er i$ and
$\fs \er{i+1}$ will be denoted by $\wm i$ or $\fs \w i{i+1}\, (\hbar = 1)$
$$\fs \wm i = \w i{i+1} = \en i - \en{i+1}, 
\qquad 1 \leq i < n. \eqno\zp$$
The second notation can be generalized as follows
$$ \w ij = \en i - \en j,\qquad 1 \leq i < j \leq n. \eqno\zp$$
The notation $\wm i$ is only defined for the energy splitting 
between successive states.

To define various atomic operators, we first introduce fermion creation 
and destruction operators $\{b_i,\ba_i\}$ for $1 \leq i \leq n$.  The
operator $\ba_i \ (b_i)$ creates (destroys) an electron in the i-th
level.  These operators satisfy the algebra
$$ \{ b_i,b_j \} = \{ \ba_i,\ba_j \} = 0, \qquad
   \{ b_i, \ba_j \} = \dl ij. \eqno\zp $$
The atomic operators can now be written as  
$$ \o ij = \ba_i b_j, \qquad 1\leq i,j\leq n \eqno\zp$$
or linear combinations of the $\o ij {\rm s}$.  The action
of $\o ij$ on an atomic state $\er k$ is given by
$$  \o ij \er k = \ba_i b_j \er k = \dl kj \er i. \eqno\zp$$
From (1.3) the general commutator for the $\cal O$ operators is
$$\bigl[ \o ij,\o kl \bigr] = \o il \dl jk - \o kj \dl il.\eqno\zp$$
Operators of the form $\o i{<j}$ are referred to as raising operators.  
These cause a transition from the lower energy
state $\er j$ to the higher energy state $\er i$.  Similarly
the operators $\o j{>i}$ are lowering operators.  We also define a set  
of commuting operators, denoted $\h m$, as follows
$$\fs \h m = \o mm - \o {m+1}{m+1}, \qquad 1\leq m\leq r. \eqno\zp$$
The set $\bigl\{\o i{\neq j},\h m\bigr\}$ satisfies (0.2), thus forming 
a representation of $\sln$.  
(Note: We shall often use the notation $X_{a<b}\ (X_{a>b})$
to mean $X_{ab}$ with $a<b\ (a>b)$ for $X$ any quantity, operator, 
c-number, etc.)

By appropriately choosing the arbitrary lowest state energy $\en n$
to be 
$$ \en n = - \sum_{m=1}^r \c{rm} \wm m, \eqno\zp$$
where $A^{-1}$ is the inverse Cartan matrix, 
the free atomic hamiltonian can be written as
$$ \ha = \sum_{1\leq u,v \leq r} \c{uv} \wm v \h u.\eqno\zp$$

\medskip

To couple the atom to an electromagnetic field we make use of the 
minimal coupling prescription.  The standard hamiltonian is
$$ H = \hp + \ha + \hi, \eqno\zp $$
where
$$ \ha = {1 \over {2 m_e}} {\p\cdot\p} + V(\x) \eqno\zp$$
$$ \hi = - {e \over {2 m_e} } \bigl( \p\cdot\ax x  + \ax x\cdot\p\, \bigr)
+ {e^2 \over {2 m}} \ax x \cdot\ax x, \eqno\zp $$
and $\hp$ is the free field hamiltonian.  If the spatial variation of
the vector potential $\a$ is small across the atom, we can take
its value at a fixed point ${\vec x_0}$ inside the atom.  Using
$$ \p = -i {m_e\over \hbar} \bigl[ \x,\ha\bigr], \eqno\zp $$
we get
$$ - {e \over {2 m_e}} \el a \p\cdot\ax x + \ax x \cdot\p\, \er b =
 {i\over \hbar} (\en b - \en a) \ax {x_0} \cdot \el a \d\, \er b, 
 \eqno\zp $$
where $\d = e\x$ is the electric dipole operator.  Since $\d$ is a
vector operator and the atomic states are assumed to be parity
eigenstates, we have $\el i \d\, \er i = 0$.  For $i<j$, the matrix 
elements are of the form 
$$ \el i \d\, \er j = d_{ij} e^{i \al ij} \n,\qquad
\el j \d\, \er i = d_{ij} e^{- i \al ij} \n,\eqno\zp $$
where $d_{ij} \geq 0$ and the unit vector $\n$ gives the spatial
orientation of the atom.  This shows that the dipole operator can be 
expanded in terms of the raising and lowering operators as
$$ \d = \n \sum_{i<j} \bigl( d_{ij} e^{i \al ij} \o ij +
d_{ij} e^{-i \al ij} \o ji \bigr).\eqno\zp $$
The interaction hamiltonian becomes
$$ \hi = -i \ax {x_0}\cdot\n \sum_{i<j} \w ij d_{ij} \bigl( 
\o ij e^{i \al ij} - \o ji e^{-i \al ij} \bigr) + {e^2 \over {2 m_e}}
\ax {x_0}\cdot\ax {x_0}.\eqno\zp$$

\medskip

To reduce this system to a one-dimensional model, we make use of the 
fibrillar geometry.  The atom can be thought of as an impurity in an 
optical fiber of cross-sectional area $\cal A$ and length $L$, with 
$L \gg {\sqrt{\cal A}}$.  Taking the fiber along the ${\widehat x}$
direction, the reduced field action is found to be (see [2] for details)
$$ {1\over \hbar} S_{\rm Maxwell} = \int dxdt\, {1\over 2}\bigl(
\dt\phi\dt\phi-\dx\phi\dx\phi \bigr), \eqno\zp$$
where $\phi$ is a dimesionless scalar field defined through
$$\a\cdot\n = \sqrt{ {4\pi \hbar}\over {{\cal A}_{\rm eff}} }\phi. \eqno\zp$$
Here $\a$ is the vector potential depending only on the $x$ coordinate and
${\cal A}_{\rm eff}$ is the effective fiber cross-sectional area.
The field $\phi$ satisfies the commutation relation
$$ [\phi(x,t),\dt\phi(x^\prime ,t) ] = i\delta(x-x^\prime).\eqno\zp$$

From the action the free field hamiltonian is found to be
$$ \hp = \int dx\, {1\over 2}\bigl[ (\dt\phi)^2 + (\dx\phi)^2 \bigr]
+ {{2 \pi e^2} \over {m_e {\cal A}_{\rm eff} }} \phi^2 (x_0),\eqno\zp $$
where the last term is the quadratic potential term taken from $\hi$.
Now the interaction Hamiltonian is
$$\hi = - {i\over 2} \phi(x_0) \sum_{i<j} \w ij \b ij \bigl(
\o ij  e^{i\al ij} - \o ji e^{-i \al ij} \bigr),\eqno\zp $$
where (explicitly showing $\hbar$ and $c$)
\zpadvnoprint
$$ \beta_{i<j} = 
\sqrt{{16 \pi} \over {\hbar c {\cal A}_{\rm eff}}} d_{ij}.
\eqno\zpstay{\rm a} $$
For $1\leq m \leq r$ we define
$$ \fs \bm m = 
\sqrt{{16 \pi} \over {\hbar c {\cal A}_{\rm eff}}} d_{m m+1} =
\b m{m+1}. \eqno\zpstay{\rm b}$$
The $\beta$ parameters are the important dimensionless coupling constants 
of the model.  The spontaneous decay rate 
$\Gamma_{ij}^s = 1/\tau_{ij}^s$ of a single excited 
atom from the state $\er i$ to the state $\er j$ is given by
$$ \Gamma_{ij}^s = {\b ij^2\over  4}\w ij. \eqno\zp $$

\medskip

Next to make the transition to a continuous system.  For $N$ atoms 
positioned at $x= x_m, m=1,\ldots,N$, let $\d_m = e(\x - \x_m)$ and
$\o ij (x_m)$ be the dipole and transition operators on the atom at
$x_m$.  The matrix elements of $\d_m$ are independent of the position,
however, the orientation can vary from atom to atom.  The operator $\d_m$
can be written in terms of the single atom matrix elements as
$$ \d_m = \n_m \sum_{i<j} d_{ij} \bigl( 
\o ij (x_m) e^{i \al ij} + \o ji (x_m) e^{-i\al ij} \bigr).\eqno\zp $$
For simplification, we consider the situation where all atoms are
aligned $\n_m = \n$ (e.g., by an external electric field), giving
$$ \hi = - {i\over 2} \int dx\, \phi(x) \sum_{i<j} \w ij \b ij 
\bigl( \o ij (x,t) e^{i \al ij} - \o ji (x,t) e^{-i\al ij} \bigr),
\eqno\zp $$
where we have introduced the space-time dependent transition operators
$$ \o ij (x,t) = \s m1N \o ij (x_m,t)\delta(x - x_m). \eqno\zp $$
The discrete operator $\o ij (x_m,t)$ acts only on the atom at $x_m$ to
cause a transition from $\er j$ to $\er i$. The $\h m (x,t)$ operators
are defined similarly.  The general commutator for 
the space-time transition operators is
$$ \bigl[\o ij (x,t),\o kl ({x^\prime },t) \bigr] = 
\bigl( \o il (x,t) \dl jk - \o kj (x,t) \dl il \bigr) 
\delta(x - {x^\prime }), \eqno\zp $$
from which it is easily seen that the algebra (or now more appropriately 
current algebra) satisfied by the set $\{ \o i{\neq j} (x,t),\h m (x,t) \}$ 
is identical, aside from the delta function factor, to
the $\sln$ algebra (0.2).  The free atomic hamiltonian takes the form
$$ \ha = \int dx\, \sum_{1\leq u,v \leq r} 
\c{uv} \wm v \h u (x,t).\eqno\zp $$

The complete hamiltonian for the system is therefore $\ha + \hp + \hi$, with
$\ha$, $\hp$ and $\hi$ given by (1.29), (1.21) and (1.26) respectively.

\advsectionnumber
\bigskip
\noindent {\bf \the\sectionnumber. Two Approximations and Integrability}
\medskip

We now make use of two approximations common in quantum optics
to further simplify $\hi$.  These being the slowly varying envelope and 
rotating wave approximations.

In the slowly varying envelope approximation, we assume that near 
resonant photons with energies $\approx \w ij$ are most relevant.  
Then the scalar field $\phi$ can be expanded about the various 
resonances as
$$ \phi(x,t) \approx \sum_{i<j}\Bigl( e^{-i \w ij (t-x)} \ps ij (x,t) +
e^{i \w ij (t-x)} \pa ij (x,t) \Bigr), \eqno\zp $$
where $\ps ij(x,t)$ and $\pa ij(x,t)$ are destruction and creation fields 
with mode expansions
\zpadvnoprint
$$ \ps i{<j} (x,0) = {1\over {\sqrt{2 \w ij}}} \int 
{dk_e \over \sqrt{2 \pi}}\, {\widehat a}_{ij} (k_e) e^{i k_e x}, 
\eqno\zpstay{\rm a} $$
$$ \pa i{<j} (x,0) = {1\over {\sqrt{2 \w ij}}} \int 
{dk_e \over \sqrt{2 \pi}}\, {\widehat a}_{ij}^\dagger (k_e) e^{-i k_e x}, 
\eqno\zpstay{\rm b} $$ 
and 
$$ {\widehat a}_{ij} (k_e) = a_{ij}(k_e + \w ij), \qquad
{\widehat a}_{ij}^\dagger (k_e) = a_{ij}^\dagger (k_e + \w ij).
\eqno\zp $$
Here $a_{ij}\ (a_{ij}^\dagger)$ is the usual photon destruction (creation) 
operator. (Note that we are only considering right-moving plane waves.)
The operator ${\widehat a}_{ij} (k_e) 
\ ({\widehat a}_{ij}^\dagger (k_e))$ destroys (creates) a photon with
energy 
$$ |k| = |k_e + \w ij| \approx \w ij. \eqno\zp $$
Thus $k_e$ acts as an ``envelope'' vector about the $\w ij$ resonance.  
The photon operators satisfy the standard commutator
$$ \bigl[ {\widehat a}_{ij} (k), 
{\widehat a}_{kl}^\dagger (k^\prime) \bigr] =
\bigl[ a_{ij} (k), a_{kl}^\dagger (k^\prime) \bigr] =
\dl ik \dl jl \delta(k - {k^\prime}). \eqno\zp $$
All other commutators vanish.  In writing (2.1) and the above commutaion 
relations, we have assumed that all resonances 
$\w ij$ ($ {1\over 2} (n^2 - n) $ in total) 
 are distinct (i.e., $\w ij = \w kl  \iff i=l\ {\rm and}\ j=k$) and sharp.  
 From (2.5) the component fields satisfy
$$ \bigl[ \ps ij (x,t), \pa kl ({x^\prime},t) \bigr] = {1\over {2 \w ij}}
\dl ik \dl jl \delta(x - {x^\prime }), \eqno\zp $$
with all other commutators zero.

The rotating wave approximation reduces the number of interactions 
in $\hi$.  Using (2.1) in $\hi$ we obtain terms with both photon creation
$(a^\dagger)$ and atomic raising ($\o i{<j}$) operators (or photon
destruction and atomic lowering $\{ a,\o j{>i}\}$), or two
photon creation (destruction) operator terms in $\hp$.  
Such high frequency terms lead to vacuum 
fluctuations and higher order processes.  The rotating wave approximation 
sets these processes to zero.  We also set to zero terms of the form
$\ps i{<j} \o k{<l} \ ({\rm or}\ \pa i{<j} \o l{>k})$ for $(i,j) \neq (k,l)$
since they give no contribution to lowest order in perturbation theory.
So we only retain those terms which pair creation/lowering or 
destruction/raising operators (and creation/destruction operators 
in $\hp$) and connect states with $\approx$ equal energy.

Combining these two approximations we get
$$ \hi = -{i\over 2} \int dx\, \sum_{i<j} \b ij \w ij 
\bigl( \ps ij e^{i \al ij} \o ij e^{-i \w ij (t-x)} - 
\pa ij e^{-i \al ij} \o ji e^{i \w ij (t-x)} \bigr),\eqno\zp $$
$$ \hp = -2 i \int dx\, \sum_{i<j} \w ij \pa ij \dx \ps ij.\eqno\zp $$
The free field hamiltonian follows from the field action, which using
$$ {|k_e|}^2 \ll \w ij^2 \Longrightarrow 
|\dx \ps ij| \ll \w ij |\ps ij|,\quad
|\dt \ps ij| \ll \w ij |\ps ij|, \eqno\zp $$
approximates to 
$$ \int dxdt\, {1\over 2} \bigl( (\dt \phi)^2 - (\dx \phi)^2 \bigr)
\approx 2 i \int dxdt\, \sum_{i<j} \w ij \pa ij (\dx + \dt) \ps ij.
\eqno\zp $$
(In (2.8) we have dropped the quadratic 
term that appears in (1.21).  For electric fields small compared with
$e/ a_0^2$, this term is negligible in relation to $\hi$.)  The phases
$e^{\pm i \al ij}$ and $e^{\pm i \w ij x}$ can be absorbed into
$\{\ps ij,\pa ij \}$ and ${\{ \o ij,\o ji \}}_{i<j}$ respectively 
without changing the commutation relations.  The time dependent
phase $e^{-i \w ij t}\ (e^{+i \w ij t})$ cancels the time dependence of
$\o i{<j}\ (\o j{>i})$  coming from the free atomic hamiltonian.  
Thus we can set  $\ha$ to zero and consider the model defined by the 
complete hamiltonian
$$ H = -2i \int dx\, \sum_{i<j} \w ij \pa ij \dx \ps ij -
{i \over 2} \int dx\, \sum_{i<j} \w ij \b ij \bigl( \ps ij \o ij -
\pa ij \o ji \bigr). \eqno\zp $$
Finally we can rescale $\ps ij$ and $\pa ij$ as
$$ \ps ij \longrightarrow {\ps ij \over \sqrt{2 \w ij} }, \qquad
\pa ij \longrightarrow {\pa ij \over \sqrt{2 \w ij} },\eqno\zp$$
which gives for the commutator (2.6)
$$ \bigl[ \ps ij (x,t), \pa kl ({x^\prime},t) \bigr] = 
\dl ik \dl jl \delta(x - {x^\prime }), \eqno\zp $$
and defining
$$\fs \wt i{<j} = {1\over 2 \sqrt{2}} \b ij \sqrt{\w ij} = 
\sqrt{\Gamma_{ij}^s \over 2}, 
\qquad \wt m{} = \wt m{m+1}\ (1\leq m\leq r),\eqno\zp $$
the hamiltonian (2.11) becomes
$$ H = -i \int dx\, \sum_{i<j} \pa ij \dx \ps ij -
i \int dx\, \sum_{i<j} \wt ij  \bigl( \ps ij \o ij -
\pa ij \o ji \bigr). \eqno\zp $$
The first interaction term, $\ps ij \o ij$, causes an atomic transition
from a lower energy state $\er j$ to a higher energy state $\er i$, along
with the absorption of a photon of energy $\approx \w ij$.  The second term, 
$\pa ij \o ji $, causes a transition from a higher energy state $\er i$ 
to a lower energy state $\er j$, along with the creation of a photon of 
energy $\approx \w ij$.

\medskip

From $H$ we can obtain the Heisenberg operator equations of motion.
Explicitly we find
\zpadvnoprint
$$\fs \eqalignno{ \ignorethis{} 
{ \dt \o k{<l} } & = 
- \hf \sum_{j>l} \wt lj \ps lj \o kj
+ \hf \sum_{i<k} \wt ik \ps ik \o il
+ \hf \sum_{i<l \atop i\not= k} \wt il \pa il \o ki & \cr
& \qquad \qquad - \hf \sum_{j>k \atop j\not= l} \wt kj \pa kj \o jl
+ \hf \wt kl \pa kl \sum_{m=k}^{l-1} \h m, & {\zpstay{\rm a}} \cr 
{ \dt \o l{>k} } & = { \dt {\o k{<l}^\dagger} } & \cr
& = - \hf \sum_{j>l} \wt lj \pa lj \o jk
+ \hf \sum_{i<k} \wt ik \pa ik \o li
+ \hf \sum_{i<l \atop i\not= k} \wt il \ps il \o ik & \cr
& \qquad \qquad - \hf \sum_{j>k \atop j\not= l} \wt kj \ps kj \o lj
+ \hf \wt kl \ps kl \sum_{m=k}^{l-1} \h m, & {\zpstay{\rm b}} \cr 
{ \dt \h m } & = 
- 2 \wt m{m+1} \bigl( \ps m{m+1} \o m{m+1} + 
\pa m{m+1} \o {m+1}m \bigr)
- \hf \sum_{j>m+1 \ignorethis{\atop j \not= l} } \wt mj  
\bigl( \ps mj \o mj + \pa mj \o jm \bigr) & \cr
& \qquad \qquad - \hf \sum_{i<m \ignorethis{\atop i \not= k}} \wt i{m+1}
\bigl( \ps i{m+1} \o i{m+1} + \pa i{m+1} \o {m+1}i \bigr) 
+ \hf \sum_{i<m} \wt im  \bigl( \ps im \o im + \pa im \o mi \bigr) 
& \cr
& \qquad \qquad \qquad + \hf \sum_{j>m+1} \wt {m+1}j 
\bigl( \ps {m+1}j \o {m+1}j + \pa {m+1}j \o j{m+1} \bigr) 
& {\zpstay{\rm c}} \cr 
} $$
$$ (\dt + \dx) \ps k{<l} = \wt kl \o lk \eqno\zpstay{\rm d} $$
$$ (\dt + \dx) \pa k{<l} = \wt kl \o kl. \eqno\zpstay{\rm e} $$
Each term in the equations of motion for the atomic operators has a simple
physical interpretation.  For example, consider the raising operator 
$\o k{<l}$.  The
first summation in (2.16a) is a sum over all atomic and photon operator pairs,
where the photon field destroys a photon of energy $\w l{<j}$, and the atomic
(raising) operator causes a transition from the lower state $\er j$ to the
higher state $\er k$.  The change in energy of the system corresponding to
the atomic/field pair being $\w kl$ (as is for every other term in (2.16a)).
If we think of a field destroying (creating) a photon of energy
$\w i{<j}$ as ``connecting'' atomic states $\er i$ to $\er j$ 
($\er j$ to $\er i$), along with $\o i{<j}\ (\o j{>i})$ connecting 
$\er j$ to $\er i$ ($\er i$ to $\er j$) and $\h m$ connecting $\er m$ to
$\er m$ and $\er {m+1}$ to $\er {m+1}$, then (2.16a) is, aside from 
c-number factors,
a sum over all atomic/field pairs connecting $\er l$ to $\er k$ through
some intermediate state, i.e., $\er l \rightarrow \er j \rightarrow \er k$.
A similar interpretation follows for $\o l{>k}$ and $\h m$.  

These equations of motion have a zero-curvature representation
\zpadvnoprint
$$ [\dt + A_t,\dx + A_x] = 0,\eqno\zpstay{\rm a} $$
$$ \Longrightarrow \dt A_x - \dx A_t = [A_x,A_t], \eqno\zpstay{\rm b} $$
where $A_x$ and $A_t$ are matrices of quantum operators given by
\zpadvnoprint
$$\eqalignno{ 
A_x = \mu \sum_{1 \leq m,n \leq r} \c {mn} \h n H^\rho_m +
\mu \sum_{i<j} \bigl( \o ij E^\rho_{ij} + \o ji E^\rho_{ji} \bigr)
& +\hf \sum_{i<j} \wt ij \bigl(-\pa ij E_{ij}^\rho + \ps ij E^\rho_{ji} \bigr)
& \cr
& - {1\over\mu}  \sum_{1 \leq m,n \leq r} 
\c {mn} \wt n{}^2 H^\rho_m 
& {\zpstay{\rm a}} \cr} $$
$$ A_t = {1\over\mu}  \sum_{1 \leq m,n \leq r} 
\c {mn} \wt n{}^2 H^\rho_m 
-\hf \sum_{i<j} \wt ij \bigl(-\pa ij E_{ij}^\rho + \ps ij E^\rho_{ji} \bigr), 
\eqno\zpstay{\rm b} $$
\underbar{provided} that the $\wt i{<j}$s satisfy
$$ \wt i{<j}^2 = \wt i{<k}^2 + \wt k{<j}^2,
\eqno\zp $$
for any intermediate value of k, $i<k<j$.
This is trivially satisfied by $\fs \wt i{i+1}$, in which case there
is no such $k$.  Here
$\c {mn}$ is the inverse Cartan matrix, and
$\{E_{i\neq j}^{\rho},H_m^{\rho}\}$ 
are matrices in any representation $\rho$ of $\sln$ satisfying (0.2). 
Requiring (2.17) to be valid for all values of 
the arbitrary spectral parameter $\mu$, is equivalent to the equations of 
motion (2.16) (of course provided (2.19) is satisfied).  The equivalence can be 
shown by making use of the commutation relations (0.2).  
From (2.16c) we can derive a more compact form for the 
equations of motion corresponding to the $\h m$ basis operators
$$ \dt \h m = - \hf \sum_{u=1}^r A_{mu} \sum_{u<l\leq n} \sum_{k=1}^u \wt kl
 \bigl( \o kl \ps kl + \o lk \pa kl \bigr). \eqno\zp $$
\ignorethis{ which gives for the general Cartan operators
$$ \dt H_{i<j} = - \hf \sum_{m=i}^{j-1} \sum_{u=1}^r A_{mu} \sum_{u<l\leq n} 
\sum_{k=1}^u \w kl
\b kl \bigl( E_{kl} \ps kl + E_{lk} \pa kl \bigr). \eqno\zp $$}

The constraint (2.19) arises in forming a zero-curvature representation for the 
field ($\{\ps ij,\pa ij\}$) equations of motion, and reduces the number of 
free parameters to $r = n-1$, these being $\{\wt m{}\}_{1\leq m\leq r}$.
The definition (2.14) shows that the constraint is equivalent to the
requirement that the spontaneous decay rate (for a single atom) 
from $\er i$ to $\er j$,
$\Gamma_{ij}^s$, be equal to the sum of the decay rates $\Gamma_{ik}^s$
and $\Gamma_{kj}^s$ for any intermediate state $\er k$
$$ \Gamma_{i<j}^s = \Gamma_{i<k}^s + \Gamma_{k<j}^s. \eqno\zp $$
The equations of motion for the atomic 
operators have the zero-curvature representation (2.17) independent of the 
constraint.  

\medskip
A zero-curvature representation implies that the system is integrable.
Thus the model (2.15) can now be solved by the Quantum Inverse 
Scattering Method.

\bigskip
\noindent {\bf References}
\medskip

\noindent
[1] V.~I.~Rupasov, JETP Lett.~36 (1982) 142.

\noindent
[2] A.~LeClair, {\it QED for a Fibrillar Medium of Two-Level Atoms},
hep-th/9604100.

\end